\documentclass[prb,twocolumn,showpacs]{revtex4}
\usepackage[dvips]{graphicx}

\newcommand{\ignore}[1]{}

\begin{document}

\title{Simple Two-Dimensional Model for the Elastic Origin of Cooperativity among Spin States of Spin-Crossover Complexes}
\author{Masamichi Nishino$^{1,4}$}
\email[Corresponding author. Email address: ]{nishino.masamichi@nims.go.jp} 
\author{Kamel Boukheddaden$^{2}$}
\author{Yusuk\'e Konishi$^{3,4}$} 
\author{Seiji Miyashita$^{3,4}$}
\affiliation{$^{1}${\it Computational Materials Science Center, National Institute}  
for Materials Science, Tsukuba, Ibaraki 305-0047, Japan \\
$^{2}$ {\it  Groupe d'Etudes de la Mati\`{e}re Condens\'{e}e, CNRS-Universit\'{e}} 
de Versailles/St. Quentin en Yvelines 45 Avenue des Etats Unis, F78035 Versailles Cedex, France \\
$^{3}${\it Department of Physics, Graduate School of Science,}
The University of Tokyo, Bunkyo-Ku, Tokyo, Japan  \\
$^{4}${\it CREST, JST, 4-1-8 Honcho Kawaguchi, Saitama, 332-0012, Japan}
}
\date{\today}

\begin{abstract}

We study the origin of the cooperative nature of spin crossover (SC) 
 between low spin (LS) and high spin (HS) states from the view point of elastic interactions among molecules. 
As the size of each molecule changes depending on its spin state, the elastic interaction among the lattice distortions provides the cooperative interaction of the spin states.
We develop a simple model of SC with intra and intermolecular potentials 
which accounts for the elastic interaction including the effect of the inhomogeneity of the spin states, and apply constant temperature molecular dynamics based on the Nos\'e-Hoover formalism. 
We demonstrate that, with increase of the strength of the intermolecular interactions, the temperature dependence of the HS component changes from a gradual crossover to a first-order transition.

\end{abstract}


\maketitle

The discovery of LIESST (light-induced excited spin state trapping) \cite{Gutlich,Decurtins} phenomena has accelerated studies of functional spin-crossover (SC) molecular solids.
SC compounds have been studied intensively not only because of their potential 
applicability to novel optical devices, e.g., optical data storage and optical sensors, etc, but also because of the fundamental scientific interest in the mechanism of the phase transition and the accompanied non-linear relaxation processes \cite{Gutlich,Decurtins,Kahn,Letard2,Hauser,Real,Shimamoto}.

To control electronic and magnetic states of SC compounds, it is important to understand the bistable nature of these compounds.
The SC transition between the low-spin (LS) and high-spin (HS) states can be induced by change of temperature, pressure, magnetic field, light-irradiation, etc.
It has been clarified that the interaction between spin states causes various types of cooperative phenomena between the LS and HS phases \cite{Sorai1}. 

In order to take into account the cooperativity in the SC transition, 
Wajnflasz and Pick (WP) proposed an Ising model \cite{Wajnflasz2}, in which the spin state is described by a fictitious spin ($\sigma=-1$ (1) for the LS (HS) state) and the short-range interactions $J$ between the spin states are introduced
in the form 
${\cal H} =- J\sum_{\langle i,j\rangle }\sigma_{i} 
\sigma_{j} + \sum_{i} (\Delta -\frac{1}{2} k_{\rm B} T \ln g ) \sigma_i$, 
where $g$ is the degeneracy ratio between the HS and LS states. 
Using this model, the change between a gradual crossover and a first-order transition has been well explained as a function of the parameters $J$, $\Delta$ and $g$.
So far the WP model and its extensions called ``Ising-like models" have been widely used for the description of 
the SC transition and related relaxation phenomena including photoinduced 
effects. 
Although the Ising-like models have captured several important features \cite{Bousseksou1,Kamel1,Nishino1,Nishino_dynamical,Miya2}, 
the origin of the parameters remains unclear due to the drastic simplifications involved.

The importance of the elastic interaction in 
the SC transition has been investigated \cite{Zimmermann,Kambara,
Adler,Willenbacher,Tchougreeff,Spiering,Nasser,kbo} 
and the elastic energy of the system with the density distribution of 
the LS and HS sites has been phenomenologically analyzed \cite{Tchougreeff}. 
The dependences of elastic constants on the spin state have been also investigated in a one-dimensional (1D) two-level model \cite{Nasser} and in a 1D vibronic coupling model \cite{kbo}. 
However, local degrees of freedom (change of lattice) can be traced out in one dimension. Thus, no phase transition occurs in one dimension.

\begin{figure}[t]
\centerline{\includegraphics[clip,width=6.0cm]{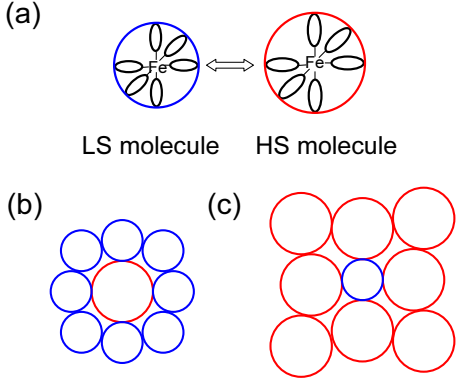} }
\caption{(a) a schematic picture of the LS molecule and the HS molecule. 
The HS molecule is bigger than the LS molecule.
(b) ((c)) shows a schematic picture of the lattice distortion for a HS (LS) molecule and surrounding LS (HS) molecules.
}
\label{Fig_distortion}
\end{figure}

Through the electron-distortion interaction, that is the vibronic coupling, 
the size of the molecule changes with the spin state. The distance (relative coordinate) between the central transition metal and the surrounding ligands changes. 
This distortion causes interactions among the spin states of molecules as depicted in Figs.~\ref{Fig_distortion} (a), (b), and (c). 
In the present study, we focus on the lattice distortions in higher dimensions caused by the difference of the molecular sizes due to the different spin states. These local distortions interact with one another elastically which causes a long range effective interaction between the spin states.

We perform molecular dynamics (MD) simulations on a 2D system with a simple square lattice. 
The intramolecular potential energy depending on the molecular size is given by a double-well adiabatic potential $V_i^{\rm intra}(r_i)$, which is a function of the radius $r_i$ of the $i$th molecule. 
Let $p_i$ be the corresponding relative momentum and let $m$ be 
the reduced mass.

\begin{figure}[thb]
\vspace*{0.8cm}
\centerline{\includegraphics[clip,width=4.6cm]{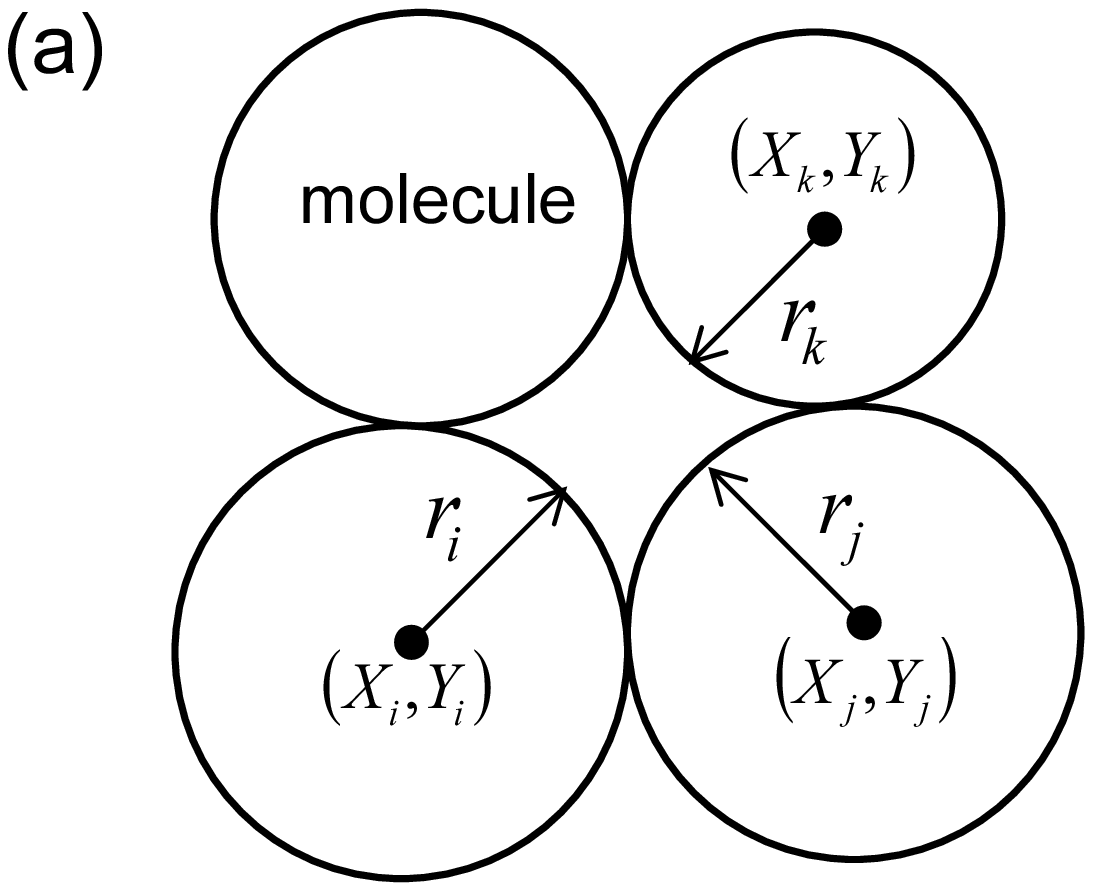} }
\vspace*{0.3cm} 
\centerline{\includegraphics[clip,width=4.5cm]{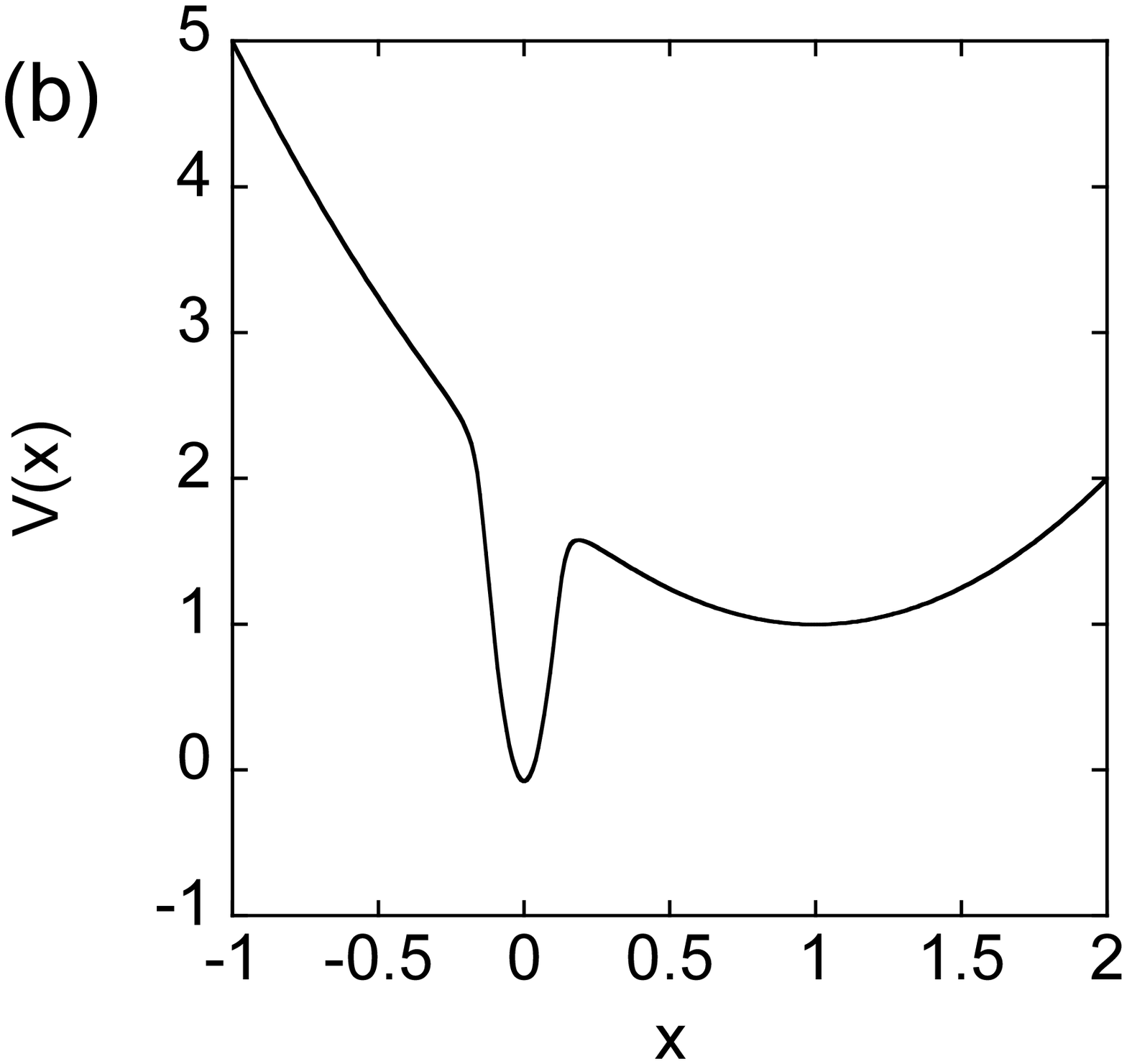} 
\includegraphics[clip,width=4.4cm]{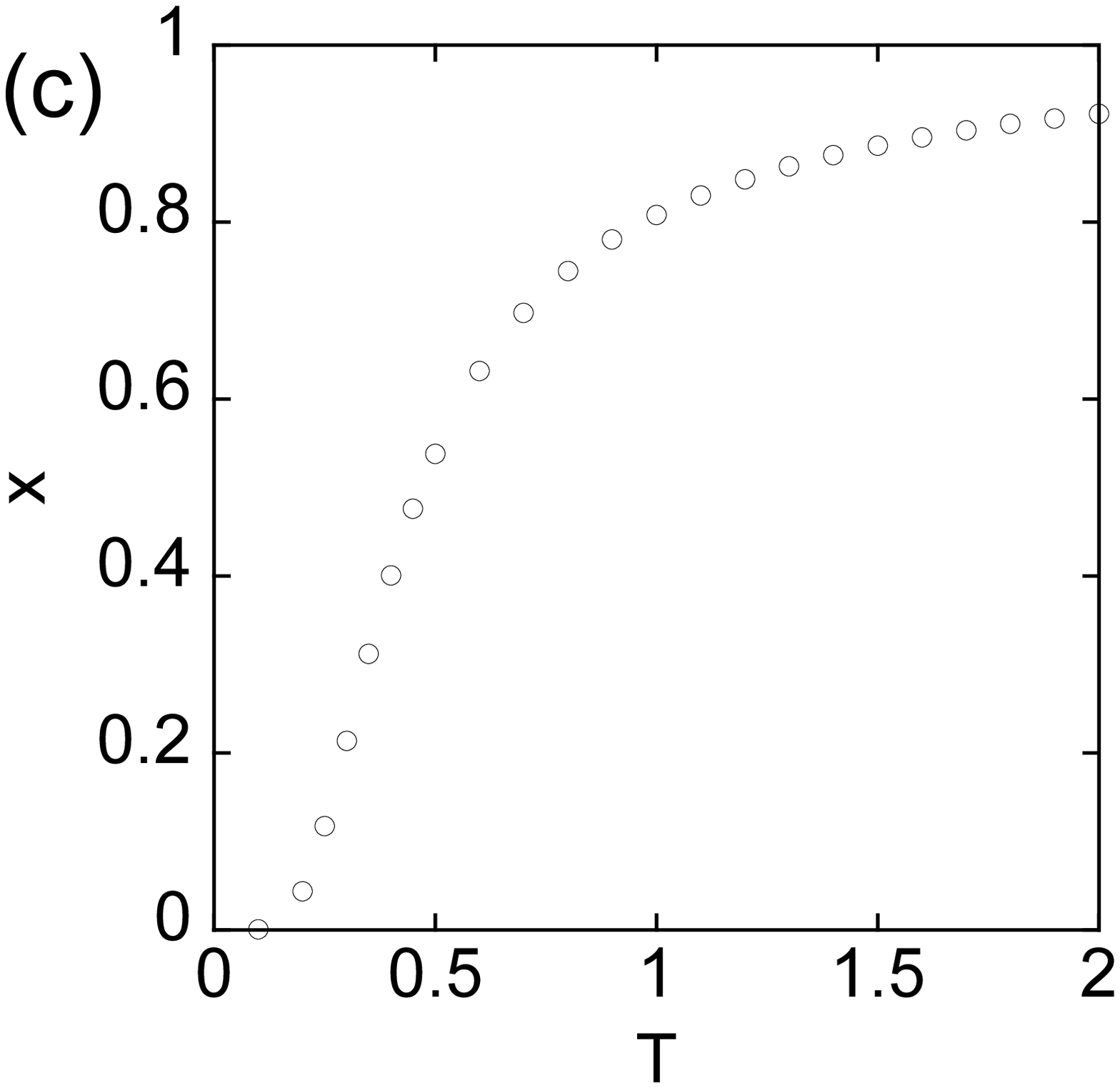}}
\caption{
(a) a schematic picture of the model.  
$(X,Y)$ is the coordinate of the center of each molecule and 
$r$ is its radius. 
(b) Intramolecular potential, where $x$ is the growth of $r$ from 
$r_{\rm LS}$.
(c) Temperature dependence of $\langle x \rangle$ without intermolecular interactions. 
}
\label{Fig_model}
\end{figure}

We set an intermolecular binding interaction between SC molecules (the $i$th and $j$th molecules) as $V_{ij}^{\rm inter}( {\mbox{ \boldmath $X$}_i}{\mbox{ \boldmath $X$}_j},r_i,r_j)$, where $\mbox{ \boldmath $X$}_i=(X_i, Y_i)$ is the coordinate of the center of the $i$th molecule (Fig.~\ref{Fig_model} (a)). 
 The corresponding momentum is $\mbox{ \boldmath $P$}_i=(P_{X_i}, P_{Y_i})$ and the mass of the molecule is $M$.

To model this scenario we apply the following Hamiltonian:  
\begin{eqnarray}
{\cal H}_{\rm system} = && \sum_{i} 
\frac{ \mbox{ \boldmath $P$}_i^2}{2 M}  + \sum_{i} \frac{p_i^2}{2 m}
+ \sum_{i} V_i^{\rm intra}(r_i)    \\ 
&& +  \sum_{\langle i,j\rangle
} V_{ij}^{\rm inter}( {\mbox{ \boldmath $X$}_i},{\mbox{ \boldmath $X$}_
j},r_i,r_j).  \nonumber
\end{eqnarray}

For simplicity we here consider only one symmetric vibration mode 
(isotropic volume expansion of molecules) as an active dominant mode \cite{Bousseksou2}.

In order to clarify the effect of distortion, 
we adopt intermolecular binding potentials independent of the molecular states, although it is expected that the intermolecular binding is looser for HS molecules than for LS molecules. 

As the intramolecular potential, we adopt a double well parabolic function $V(x)$, where $x$ is defined as the difference of the radius 
from that of the ideal LS state. $V(x)$ has minima at $x=0$ (ideal LS) and $x=1$ (ideal HS).
Setting $r_{\rm LS}=9$ and $r_{\rm HS}=10$ for the ideal radius of the LS molecule and that of the HS molecule, respectively, the radius of the 
molecule is $r=r_{\rm LS}+x$. 

When a parabolic potential for the LS state 
($y=ax^2$) and that for the HS state ($y=b(x-c)^2+d$) are mixed by off-diagonal element $J$, 
the lowest potential function with coefficient $A$ is given by 

\begin{eqnarray}
V(x)= &&  \frac{A}{2}  \{ d+b(c-x)^2 + a x^2     \\
      && -\sqrt{4 J^2+(d+b(c-x)^2-ax^2)^2}  \}. \nonumber    
\end{eqnarray}
Because the entropy of the harmonic oscillator (${\cal H}=\frac{p^2}{2m}+\frac{1}{2}K x^2$) is $S=Nk_{\rm B} \left( 1-\ln 
\frac{\hbar}{k_{\rm B} T} \sqrt{\frac{K}{m}} \right)$, the entropy difference between the LS and HS states is given by 
\begin{equation}
\Delta S =S_{\rm HS}-S_{\rm LS}=Nk_{\rm B} \ln 
\sqrt{  \frac{K_{\rm LS}}{K_{\rm HS}}}.
\end{equation} 
Thus the ratio of the degeneracy between the HS and LS states 
is $g=\sqrt{  \frac{K_{\rm LS}}{K_{\rm HS}}}=\sqrt{\frac{a}{b}}$, which corresponds to 
$g$ in the WP-model. 
We take $A = 10$, $a=10$, $b=0.1$, $c=1.0$, $d=0.1$, $J=0.04$,  
which gives $g=10$. In realistic materials there are several sources of the difference of the entropy. 
However, in order to 
focus on only the lattice effect, we adopt a large value of $\sqrt{  \frac{K_{\rm LS}}{K_{\rm HS}}}(=10)$.
In Fig.\ref{Fig_model} (b), the intramolecular potential $V(x)$ is depicted.
The energy difference between the LS stable point and the HS stable point is $\Delta E_{\rm LS-HS} = 1.075$ and the energy difference between the LS stable point and the unstable point ($x_{\rm c}=0.19$) is $\Delta E_{\rm act.}=1.654$. 
The temperature dependence of the statistical average of $x$, i.e., 
\begin{equation} 
\langle x \rangle = \frac{{\rm Tr} \: x \exp( -\beta {\cal H}_{\rm system} )}{{\rm Tr} \exp( -\beta {\cal H}_{\rm system})}
\end{equation} 
for $V_{ij}^{\rm inter}=0$, calculated by a numerical integration, is given in 
Fig.~\ref{Fig_model} (c).

Next, we consider the intermolecular potential: $ V_{ij}^{\rm inter}( {\mbox{ \boldmath $X$}_i},{\mbox{ \boldmath $X$}_
j},r_i,r_j)$ between the nearest neighbors ($i$th and $j$th molecules).
We take
\begin{eqnarray}
V_{ij}^{\rm inter}( {\mbox{ \boldmath $X$}_i},{\mbox{ \boldmath $X$}_
j},r_i,r_j) = && f(d_{ij}),
\label{NNpair}
\end{eqnarray}
where $d_{ij}=| {\mbox{ \boldmath $X$}_i} - {\mbox{ \boldmath $X$}_j} |- (r_i+r_j)$ .
We treat phenomena in which the lattice distortion is not 
so large and it does not break the lattice structure. 
In this case, the qualitative 
features of the phenomena do not depend on the details of the potential form, 
and thus we adopt one of the simplest forms  
\begin{eqnarray}
f(u) = && D \left( e^{a' (u-u_0)} + e^{-b' (u-u_0)} \right), 
\end{eqnarray}
 where $a'=0.5$ and $b'=1.0$.
$u_0$ is a constant such that $f(u)$ has the minimum at $u=0$. 
When molecules (circles in Fig.~\ref{Fig_model} (a)) $i$ and $j$ contact each other ($d_{ij}=0$), the function has a minimum value.
 
 In order to maintain the crystal structure (the coordination number), 
we introduce a potential between the next-nearest neighbors ($i$ and $k$, see Fig.~\ref{Fig_model} (a))
\begin{eqnarray}
V_{ik}^{\rm inter}( {\mbox{ \boldmath $X$}_i},{\mbox{ \boldmath $X$}_
k},r_i,r_k) = && f(d_{ik}-\Delta r) 
\label{NNNpair}
\end{eqnarray}
with $a'=0.1$ and $b'=0.2$, which is much smaller than that of the nearest neighbors.
Next-nearest neighbors do not contact each other as depicted in 
Fig.~\ref{Fig_model} (a), and there is a spatial gap between them.
For simplicity, we assume here that next-nearest neighbors are most stabilized when 
the gap is $\Delta r= 2(\sqrt{2}-1) \bar{r}$, 
where we take $\bar{r}=(r_{\rm LS} + r_{\rm HS}) /2$ although $\Delta r$ can be temperature dependent. 
We focus on the dependence of the spin state on 
the strength of the intermolecular interaction, and thus we study the dependence on $D$. 
Common $D$ is used for both Eqs.~(\ref{NNpair}) and (\ref{NNNpair}).

To study the temperature dependence, 
we adopt the Nos\'e-Hoover method \cite{Nose,Hoover} to generate the canonical ensemble for a given temperature $T$. 
The Hamiltonian of the thermal reservoir is given by 
\begin{equation}
{\cal H}_{\rm therm} = \frac{P_s^2}{2 Q}  + 3N k_{\rm B} T \ln s,
\end{equation}
where $s$ is a scaling factor, $P_s$ is the conjugated momentum of $s$ and $Q$ 
is an effective mass associated with $s$. 
Therefore, the total Hamiltonian including the effect of thermal reservoir is 
given by ${\cal H}_{\rm total} = {\cal H}_{\rm system} + {\cal H}_{\rm therm}$.

Applying the Nos\'e-Hoover formalism to the present system, 
the time evolution of the system is realized according to the following equations of motion.
\begin{eqnarray}
 \frac{dr_i}{dt}  &=&  \frac{p_i}{m}, \\
 \frac{dp_i}{dt}  &=&  -\frac{\partial V^{\rm intra}}{\partial r_i}
 -\frac{\partial V^{\rm inter}}{\partial r_i} - \xi p_i, \\
  \frac{d\mbox{\boldmath $X$}_i}{dt}  &=& \frac{\mbox{\boldmath $P$}_i}{M}, \\
 \frac{d\mbox{\boldmath $P$}_i}{dt}  &=&  -\frac{\partial V^{\rm inter}}{\partial \mbox{ \boldmath $X$}_i}  - \xi \mbox{\boldmath $P$}_i, \\
 \frac{ds}{dt}  &=&  s \xi, \\
 \frac{d\xi}{dt}  &=&  \frac{1}{Q} \left[\sum_i \frac{p_i^2}{m}+ \sum_i 
 \frac{P_i^2}{M}  -3N k_{\rm B} T  \right] , 
\end{eqnarray}
where $V^{\rm inter}$ stands for the summation of the intermolecular potentials for
the nearest and next-nearest pairs, and $\xi \equiv \frac{P_s}{Q}$.

We adopt $x$ as a parameter to characterize the spin state. 
We study the open-boundary system of $L^2=26 \times 26$ molecules.
We warm up the system from $T=0.1$ to $2.0$ in steps of increment 0.1, 
and cool it down to $T=0.1$. 
At each temperature, 40000 MD steps are discarded as transient time 
and subsequent 20000 MD steps are used to measure $x$ with the time step $\Delta t=0.01$.
We employ an operator decomposition method in which 
the numerical error is of the order $O(\Delta t^3)$.
We set $m=1.0$, $M=1.0$, and $Q=1.0$. ($\langle x \rangle$ does not depend on $m$, $M$, and $Q$ in the equilibrium state.)

\begin{figure}[t]
\centerline{\includegraphics[clip,width=4.5cm]{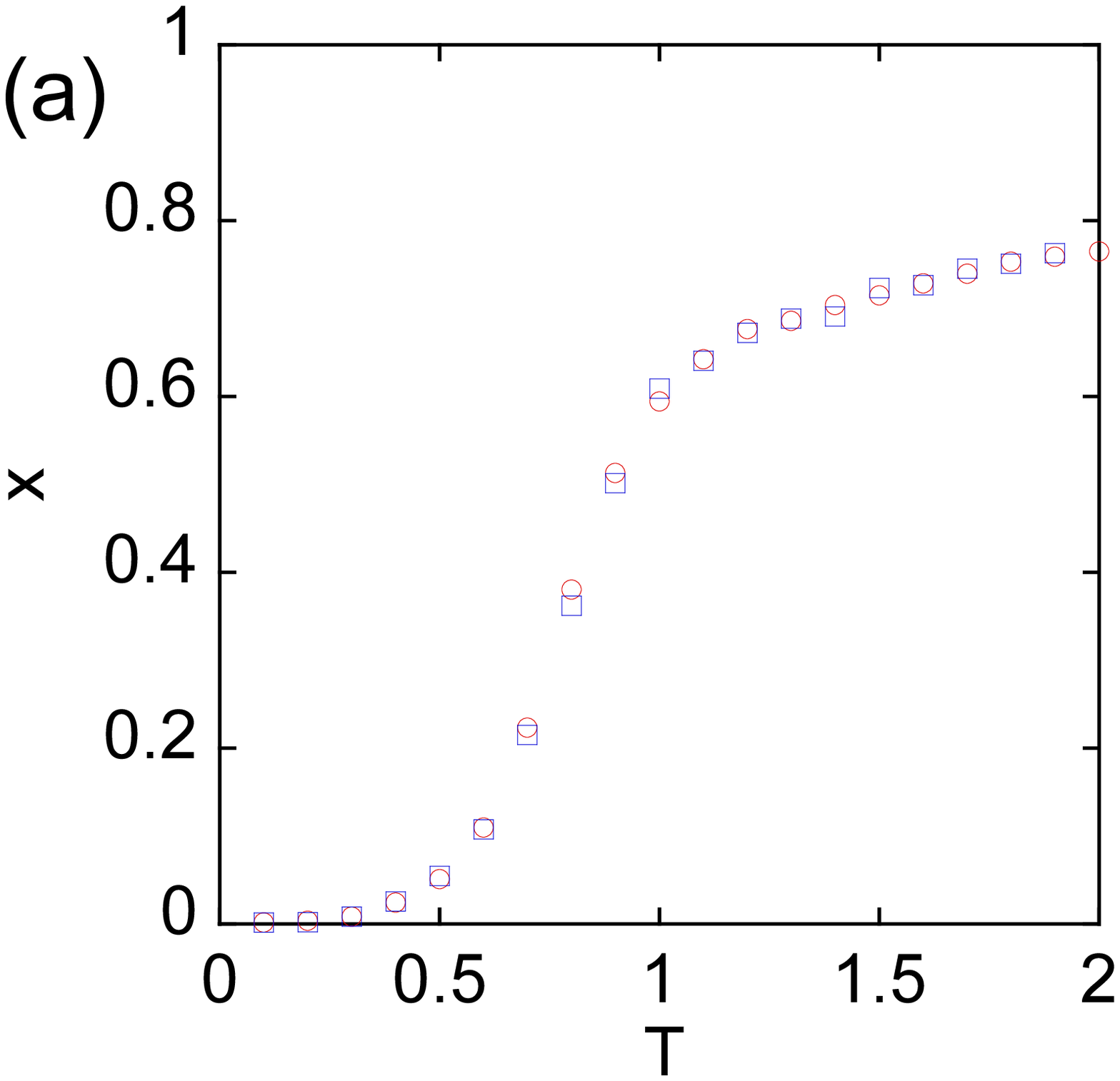} \includegraphics[clip,width=4.5cm]{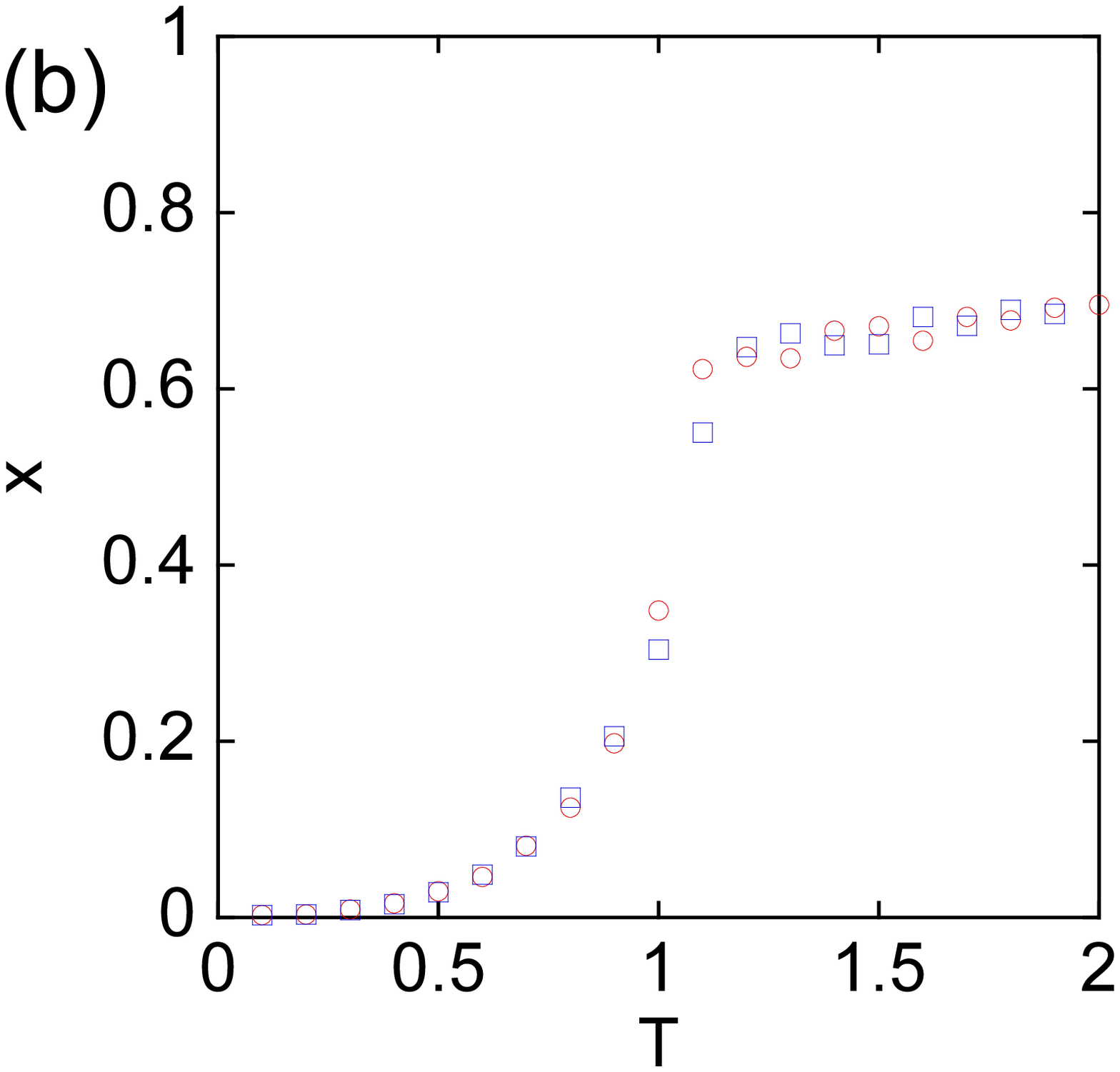}}
\vspace*{0.3cm}
\centerline{\includegraphics[clip,width=4.5cm]{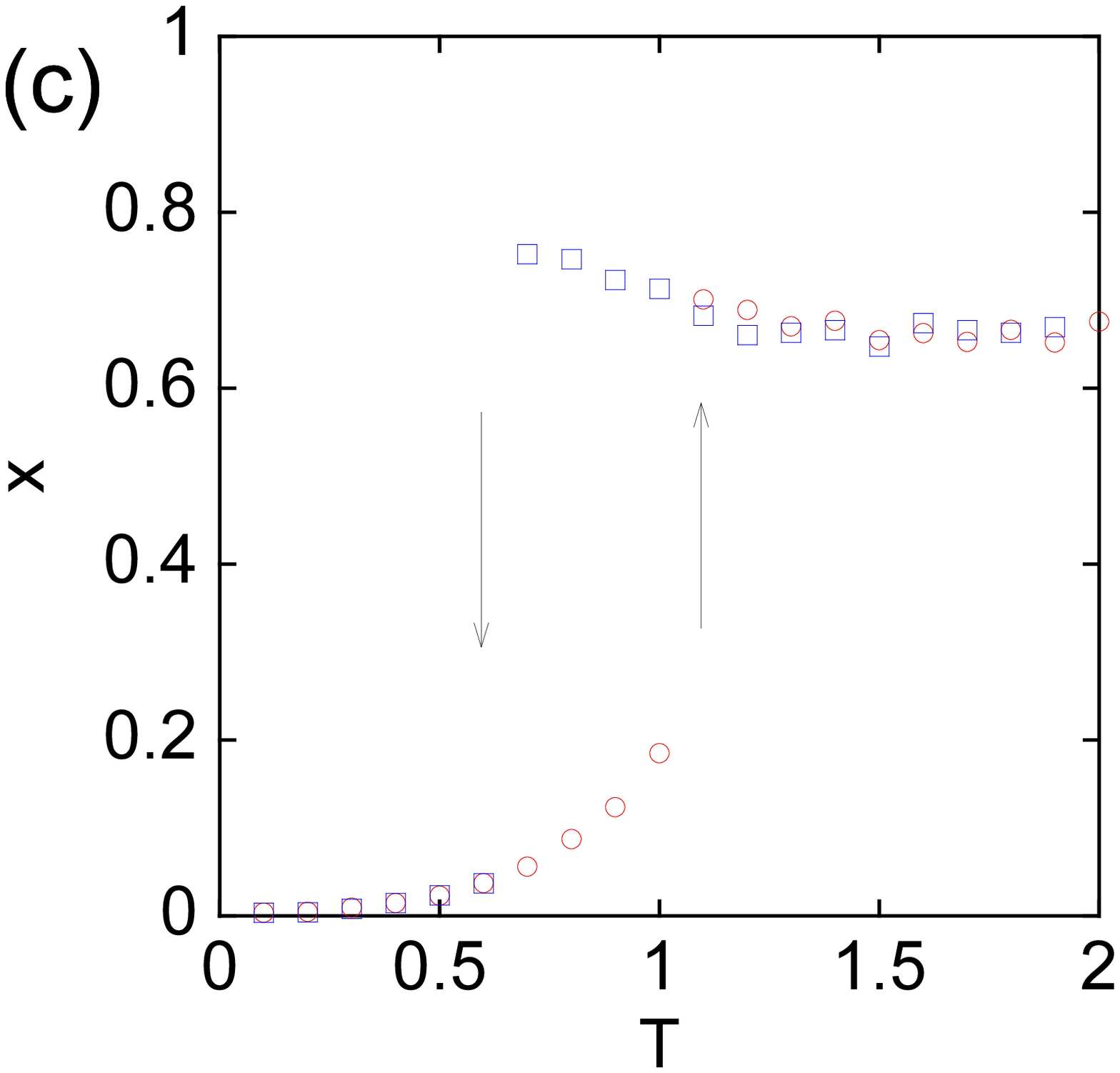} \includegraphics[clip,width=4.5cm]{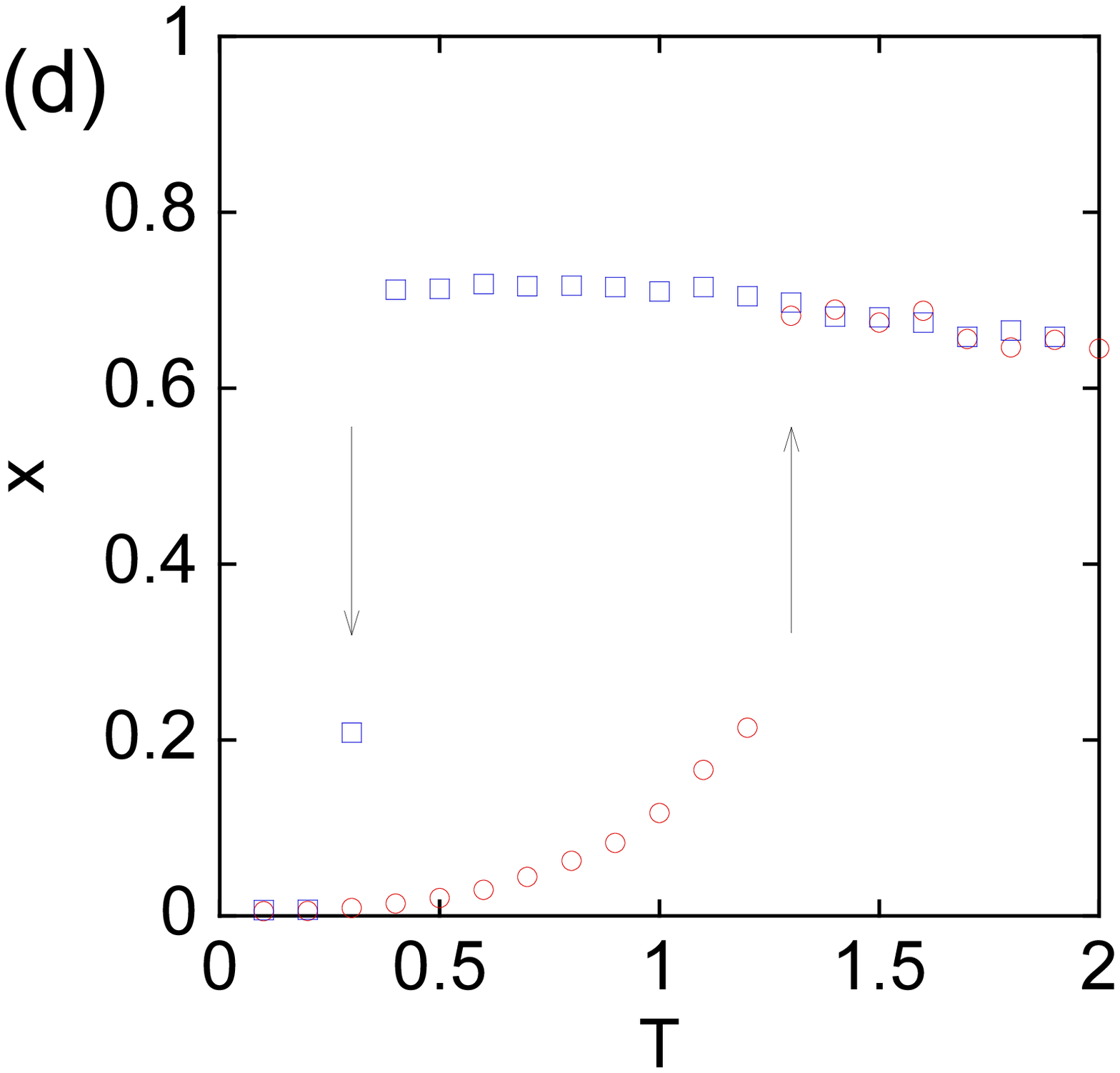}}
\caption{Temperature dependences of $\langle x \rangle$ for the values of (a) $D=10$, (b) $D=20$, (c) $D=28$, and (d) $D=42$.
The open red circles (blue squares) denote $\langle x \rangle$ in the warming (cooling) process. 
}
\label{Fig_T_x}
\end{figure}

In Figs.~\ref{Fig_T_x} (a), (b), (c) and (d), the temperature dependences 
of $\langle x \rangle $ are shown. 
When $D=10$ (Fig.~\ref{Fig_T_x} (a)), $\langle x \rangle$ in the warming process and that in the cooling process overlap, indicating that 
a smooth (gradual) SC crossover is realized. 
When the interaction parameter becomes larger: $D=20$ (Fig.~\ref{Fig_T_x} (b)),
variation of $\langle x \rangle$ becomes sharper, which implies that the SC transition becomes more cooperative.

When $D=28$ (Fig.~\ref{Fig_T_x} (c)), a clear hysteresis loop of $\langle x \rangle$ is found. 
As the interaction parameter increases further: $D=42$ (Fig.~\ref{Fig_T_x} (d)), the hysteresis width becomes larger. 
Here, we found that when the interaction between molecules becomes large, 
the SC transition changes from a gradual crossover to a first order transition. 
The critical value of $D$ is $D_{\rm critical} \simeq 20$.

\begin{figure}[t]
\centerline{\includegraphics[clip,width=6.5cm]{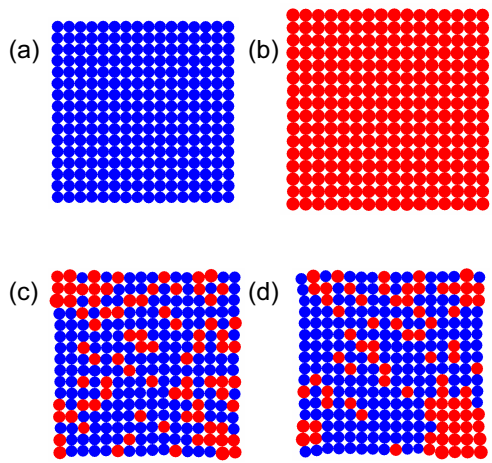}}
\caption{Snapshots of configurations. 
HS molecules (red circles) are allocated when $r$ is larger than $r=r_{\rm LS}+x_{\rm c}$. LS molecules are drawn by blue circles. 
(a) Complete LS phase. 
(b) Complete HS phase.
(c) A snapshot of configuration at $T=0.6$ in the system of $D=10$ ($L^2=16^2$), where 80 molecules are in the HS state.
(d) A snapshot of configuration at $T=1.0$ in the system of $D=42$ ($L^2=16^2$), where 79 molecules are in the HS state.
}
\label{snapshot}
\end{figure}

In Figs.~\ref{snapshot} (a) and (b), snapshots of the complete LS state and the complete HS state are shown, in which the system length changes by 11 \%.
In Figs.~\ref{snapshot} (c) and (d), a snapshot of configuration at $T=0.6$ 
for the parameter $D=10$ and that at $T=1.0$ for $D=42$ are given, where the concentration of HS molecules is about 30\% in both configurations.  
Although the number of HS molecules is almost the same, 
the average cluster size of HS in (d) is bigger than in (c). 
This indicates that there is higher correlation between spin states of molecules in the case of strong intermolecular interaction (case (d)), which promotes first-order transition. 

In this study, we investigated the cooperativity of spin-crossover phenomena induced by the elastic interaction among lattice distortions which are triggered by the difference of molecular sizes caused by the different spin states. 
This effect is inherent to the high dimensionality (2D and 3D). The present 2D model can be applied straightforwardly to the 3D case. 
Although the lattice relaxation through a change of molecular sizes has been studied phenomenologically by a mean-field treatment \cite{Tchougreeff}, as far as we know, this is the first attempt to investigate the cooperativity attributed to the effect of local distortions (fluctuation) and that of the propagation to the overall lattice.

The present work was supported by Grant-in-Aid for Scientific Research on 
Priority Areas and for Young Scientists (B) from MEXT of Japan, 
and supported by Grant-in-Aid from Minist\`{e}re de l'Education Nationale and CNRS (PICS France-Japan Program) of France. 
The present work was also supported by the MST Foundation. 
The numerical calculations were supported by the supercomputer center of
ISSP of Tokyo University.

\end{document}